\documentclass[twocolumn,nofootinbib,prd,superscriptaddress]{revtex4}
\usepackage{xcolor}
\usepackage{amsmath}
\usepackage{xspace}
\usepackage[normalem]{ulem}
\usepackage{graphicx}
\usepackage{hyperref}
\newcommand\skipme[1]{}
\newcommand\unit[1]{{\rm #1}}

\newcommand\mysec[1]{\section{#1}}

\newcommand\qmstateproduct[2]{\left\langle#1|#2\right\rangle}

\newcommand{\IMRP}{\textsc{IMRPhenomPv2}\xspace}
\newcommand{\SEOB}{\textsc{SEOBNRv3}\xspace}
\def\RIT{Center for Computational Relativity and Gravitation, Rochester Institute of Technology, Rochester, New York 14623, USA}
\def\GT{Center for Relativistic Astrophysics and School of Physics, Georgia Institute of Technology, Atlanta, Georgia 30332, USA}
\def\GLA{SUPA, University of Glasgow, Glasgow G12 8QQ, United Kingdom}
\def\CITA{Canadian Institute for Theoretical Astrophysics, University of Toronto, Toronto, Ontario M5S 3H8, Canada}
\def\COR{Cornell Center for Astrophysics and Planetary Science, Cornell University, Ithaca, New York 14853, USA}
\def\RAD{Department of Astrophysics/IMAPP, Radboud University, Nijmegen, P.O. Box 9010, 6500 GL Nijmegen, The Netherlands}

\begin{document}

\title{Systematic challenges for future gravitational wave measurements of precessing binary black holes\\}
\author{A.~R.~Williamson}
\affiliation{\RIT}
\affiliation{\RAD}
\author{J.~Lange}
\affiliation{\RIT}
\author{R.~O'Shaughnessy}
\affiliation{\RIT}
\author{J.~A.~Clark}
\affiliation{\GT}
\author{Prayush~Kumar}
\affiliation{\CITA}
\affiliation{\COR}
\author{J.~Calder\'{o}n~Bustillo}
\affiliation{\GT}
\author{J.~Veitch}
\affiliation{\GLA}
\begin{abstract}
    The properties of precessing, coalescing binary black holes are presently
    inferred through comparison with two approximate models of compact binary
    coalescence. In this work we show these two models often disagree
    substantially when binaries have modestly large spins ($a\gtrsim 0.4$)
    and modest mass ratios ($q\gtrsim 2$). We demonstrate these disagreements
    using standard figures of merit and the parameters inferred for recent
    detections of binary black holes. By comparing to numerical relativity, we
    confirm these disagreements reflect systematic errors. We provide concrete
    examples to demonstrate that these systematic errors can significantly
    impact inferences about astrophysically significant binary parameters. For
    the immediate future, parameter inference for binary black holes should be
    performed with multiple models (including numerical relativity), and
    carefully validated by performing inference under controlled circumstances
    with similar synthetic events.
\end{abstract}
\maketitle
\begin{table*}
\begin{ruledtabular}
\begin{tabular}{lrrrrrrrr}
NR ID/Approximant & $q$ ($m_1/m_2$) & $M_{\rm tot}$ ($M_{\rm Sun}$) & $\chi_{\rm 1x}$ & $\chi_{\rm 1y}$ & $\chi_{\rm 1z}$ & $\chi_{\rm 2x}$ & $\chi_{\rm 2y}$ & $\chi_{\rm 2z}$\\ \hline
\SEOB & 1.91 & 60.0 & -0.400 & 0.552 & -0.346 & 0.174 & -0.079 & -0.052\\
\SEOB & 3.01 & 26.5 & 0.951 & -0.115 & 0.124 & 0.510 & 0.298 & 0.760\\
SXS:BBH:0165 & 6.00 & 80.0 & -0.058 & 0.776 & -0.470 & 0.076 & -0.172 & -0.234\\
SXS:BBH:0112 & 5.00 & 80.0 & 0 & 0 & 0 & 0 & 0 & 0\\
\end{tabular}
\caption{\label{tab:SourceParameters}\textbf{Parameters of synthetic sources}:  This table shows the parameters of all the synthetic sources (Approximant and NR) used in this paper.  $q$ is the mass ratio defined with $q>1$, $M_{\rm tot}$ is the total mass, and $\chi_{*}$ are the components of the normalized spins.}
\end{ruledtabular}
\end{table*}

\mysec{Introduction}
The Laser Interferometer Gravitational Wave Observatory (LIGO) has reported the  discovery of three binary black hole
(BBH) mergers to date -- GW150914~\cite{2016PhRvL.116f1102A}, GW151226~\cite{2016PhRvL.116x1103A}, and
GW170104~\cite{2017PhRvL.118v1101A} --  along with one astrophysically plausible candidate signal, LVT151012~\cite{LIGO-O1-BBH}.
At this early stage, observations cannot firmly distinguish between a number of possible BBH formation mechanisms~\cite{AstroPaper}.
These include the evolution of isolated pairs of
stars~\cite{gwastro-EventPopsynPaper-2016,popsyn-LowMetallicityImpact-Chris2010,popsyn-LIGO-SFR-2008,2012ApJ...759...52D,2016MNRAS.458.2634M,popsyn-ChemHomogeneous-Marchant2016},
dynamic binary formation in dense clusters~\cite{2016ApJ...824L...8R}, and pairs of primordial black
holes~\cite{2016PhRvL.116t1301B}; see, e.g.,~\cite{AstroPaper} and references therein.  
One way to possibly distinguish between isolated and dynamic formation mechanisms is to measure the spin properties of
the black holes~\cite{richard-pro-forma-IsotropicVsAlignedCite,AstroPaper,gwastro-popsynVclusters-Rodriguez2016,gwastro-PE-Salvo-EvidenceForAlignment-2015,gwastro-AlignmentIdnetify-ToyModel-Stevenson2017,2017arXiv170403879O, 2017arXiv170408370T}.
The presence of a component of the black hole spins in the plane of the orbit leads to precession of that plane.  If suitably massive and significantly spinning, such binaries will strongly precess within the LIGO sensitive band.
If BBHs are the end points of isolated binary star systems, they would be expected to contain black holes with spins
preferentially aligned with the orbital angular momentum~\cite{2000ApJ...541..319K,gwastro-ConstrainChannels-BoxingDayKicks-Me2017}, and therefore rarely  be strongly precessing.
If, however, BBHs predominantly form as a result of gravitational interactions inside dense populations of stellar systems, the relative orientations of the black hole spins with their orbits will be random, and some gravitational wave signals may be very strongly precessing.
Precise measurements of their properties will provide unique clues into how black holes and massive stars evolve~\cite{2017CQGra..34cLT01V,2017arXiv170306873S,gwastro-popsynVclusters-Rodriguez2016,2017arXiv170601385F,gwastro-ConstrainChannels-KickRatePaper-2017,2016ApJ...830L..18B,2016PhRvD..94f4020N}.

The gravitational wave signals produced by strongly precessing systems are challenging to model.
Direct numerical simulations of Einstein's equations are possible for these and other generic orbits, but are time consuming to produce.
As a result, approximate models to numerical relativity have been developed, and recent models feature ways to mimic
signals from precessing systems~\cite{2014PhRvD..89f1502T,2014PhRvD..89h4006P,nr-Jena-nonspinning-templates2007,gwastro-nr-Phenom-Lucia2010,gwastro-mergers-IMRPhenomP,2015PhRvL.115l1102B,gw-astro-EOBspin-Tarrachini2012}.
These approximate models have been used to infer the properties of observed systems~\cite{PEPaper,SEOBv3Paper}.
In this paper we demonstrate by example several systematic issues which can complicate the interpretation of
rapidly-spinning and precessing binaries.
First, we  provide one of the first  systematic head-to-head comparisons of these models for precessing, coalescing
binaries, using physically equivalent parameters for both waveforms; see also~\cite{2017PhRvD..95j4023B,2017arXiv170507089B}.
We show that the two models disagree frequently for precessing systems, including parameters within the posterior distributions of
gravitational wave events like GW151226 and GW170104.  
Our study differs from several  previous investigations of waveform fidelity~\cite{2013CQGra..31b5012H,2015PhRvD..92j2001K,gwastro-mergers-nr-PrayushAlignedSpinAccuracy-2016} by focusing on precessing systems and
observationally-motivated parameters. 
The two models principally disagree when the spins are both large and precessing.   GW measurements like LIGO's have not strongly
prescribed whether such strongly-precessing systems are consistent with any individual observation.   Using concrete
examples, we remind the reader that the posterior distributions for BH spins can depend significantly on the assumed
prior distributions, particularly since these distributions are often broad and nongaussian
\citep[see, e.g.][]{2015ApJ...798L..17C,2016ApJ...825..116F,2017arXiv170704637V}.
One astrophysically plausible prior distribution is significant BH natal spin (e.g., as motivated by some X-ray
observations) and random BH spin-orbit alignment (e.g., as implied by dynamical formation scenarios).  
We show that, if these prior assumptions are adopted, the posterior distribution is dominated by parameters for which
the models  disagree even more frequently.  

We perform parameter estimation on synthetic signals to demonstrate quantitatively that these disagreements lead to biases in, and different conclusions about, astrophysically relevant quantities.
These synthetic signals have parameters and detector configurations consistent with observed events.
Extending the study of~\cite{LIGO-O1-PENR-Systematics}, which focused on weakly precessing systems, we show that
inferences about GW sources derived using the conventional configuration can frequently be biased, particularly in
certain regions of the parameter space and about observationally-relevant pairs of parameters.
We show that the conclusions reached can be strongly dependent on the model used.  
We point out that  extensive followup studies --  using multiple models and numerical relativity
-- were performed on GW150914~\cite{PEPaper,NRPaper,LIGO-O1-PENR-Systematics} and  GW170104~\cite{2017PhRvL.118v1101A,NRFollowupPaperGW170104},
producing good agreement across multiple independent calculations.

This paper is organized as follows.  In Section \ref{sec:ModelModelMismatch} we compare the predictions of two models
for the radiation emitted by graivtational waves from precessing binary binary black holes.  To make our discussion
extremely concrete and observationally relevant, we perform these comparisons on parameters drawn from LIGO's
 inferences about  GW151226, and from our inferences about  synthetic events designed to mimic GW170104 and GW151226.    Under the conventional assumptions used in this analysis, we
 find the two models disagree, principally when the inferred binary parameters involve large precessing spins.  Because
 the relative probability of large and precessing spins depends on our prior assumptions, we then repeat these
 comparison again, adopting  prior assumptions that do not disfavor BH binaries with two significant, precessing spins.
To illustrate the implications of these disagreements, in Section \ref{sec:PE:Examples}, we perform several
proof-of-concept parameter inference calculations using synthetic gravitational wave signals.  Again, using parameters
consistent with real observations (i.e., drawn from observed posterior distributions of observed BH-BH binaries), we show
that parameter inferences performed with the two models can disagree substantially about astrophysically relevant
correlated parameters, like the mass and spin of the most massive BH.    
To highlight the fact that these disagreements occur frequently, not merely for systems viewed in rare edge-on lines of sight, we choose synthetic
binaries which are inclined by $\pi/4$ to the line of sight. 
In Section \ref{sec:Discussion}, we discuss how our results extend the broadening appreciation of potential sources of systematic error in
gravitational wave measurements.

\section{Models for compact binary coalescence disagree}
\label{sec:ModelModelMismatch}

\subsection{Models for radiation from binary black holes}
When inferring properties of coalescing binary black holes~\cite{PEPaper,SEOBv3Paper,LIGO-O1-BBH,2017PhRvL.118v1101A}, LIGO has so far favored two approximate models for their gravitational
radiation: an effective one body (EOB) model, denoted \SEOB~\cite{2014PhRvD..89f1502T,2014PhRvD..89h4006P}, and a
phenomenological frequency-domain inspiral and merger model,
denoted \IMRP~\cite{gwastro-mergers-IMRPhenomP}.

The \SEOB model extends a long, incremental tradition to modeling the inspiral and spin dynamics
of coalescing binaries via an ansatz for the two-body Hamiltonian~\cite{gw-astro-EOBspin-Tarrachini2012}.  In this
approach, equations of motion for the BH locations and spins are evolved in the time domain.  For nonprecessing
binaries, outgoing gravitational
radiation during the inspiral phase is generated using an ansatz for resumming the post-Newtonian expressions for
outgoing radiation including non-quasicircular corrections, for the leading-order $\ell=2$ subspace.  For the  merger phase of nonprecessing binaries,  the
gravitational radiation is generated via a resummation of many quasinormal modes, with coefficients chosen to ensure smoothness.
The final black hole's mass and spin, as well as some parameters in the nonprecessing inspiral model, are generated via calibration to numerical relativity simulations of black hole
coalescence.   For precessing binaries, building off the post-Newtonian ansatz of seperation of timescales and orbit
averaging~\cite{1996PhRvD..54.4813W,1995PhRvD..52..821K,ACST,BCV:PTF}, gravitational radiation during the insprial is modeled as if from an  instantaneously nonprecessing
binary (with suitable nonprecessing spins), in a frame in which the binary is not precessing~\cite{2011PhRvD..84b4046S,gwastro-mergers-nr-Alignment-ROS-CorotatingWaveforms,2013PhRvD..87j4006B}.
During the merger, the radiation is approximated using the same final black hole state, with the same precession
frequency.\footnote{This choice of merger phase behavior is known to be inconsistent with precessional dynamics during merger~\cite{2013PhRvD..87d4038O,2017arXiv170507089B}.}  
With well-specified initial data in the time domain, this method can be  directly compared to  the trajectories~\cite{2015PhRvD..92j4028O}
 and radiation~\cite{2017PhRvD..95b4010B}  of numerical binary black hole spacetimes. 

The \IMRP model is a part of an approach that attempts to  approximate the leading-order gravitational wave radiation using phenomenological fits
to the fourier transform of this radiation, computed from numerical relativity simulations and post-newtonian calculation~\cite{nr-Jena-nonspinning-templates2007,gwastro-nr-Phenom-Lucia2010,gwastro-mergers-IMRPhenomP}.  Also using information
about the final black hole state, this phenomenological frequency-domain approach matches standard approximations for
the post-Newtonian gravitational wave phase to an approximate, theoretically-motivated spectrum characterizing merger
and ringdown. 
Precession is also incorporated by a ``corotating frame'' ansatz, here implemented via a stationary-phase approximation
to the time-domain rotation operations performed for \SEOB.  

We make  use of the  \texttt{lalsimulation} implementations of these two
approximations, provided and maintained by their authors in the same form as used in LIGO's O1 and O2 investigtations.

The coalescence time and orientation (i.e., Euler angles) of a binary are irrelevant for the inference of instrinic parameters from gravitational wave data.
As a result, and following custom in stationary-phase calculations, the \IMRP model does not calibrate the reference phases and time.
This makes easy head-to-head comparison with time-domain calculations somewhat more difficult.
Specifically, two different sets of parameters are needed to generate the same gravitational radiation in \SEOB and \IMRP, connected by (a) a
change in the overall orbital phase; (b) a change in the precession phase of the orbital angular momentum; and (c) a
change in the overall coalescence time.  In the approximations adopted by \IMRP, these time and phase shifts do not
qualitatively change the underlying binary or its overall orientation-dependent emission, just our perspective on it.

\subsection{Binary black hole observations and model-based inference}

Inferences about black hole parameters are performed using tools that apply Bayes' theorem and standard  Monte Carlo
inference techniques; see, e.g.,~\cite{PEPaper,gw-astro-PE-lalinference-v1} and references therein.  
For any coalescing black hole binary, fully characterized by parameters $x$, we can compute the (Gaussian) likelihood function $p(d|x)$ for any stretch of
detector network data $d$ that the signal is present in by using waveform models and an estimate of the (approximately Gaussian) detector noise on short timescales (see, e.g.,~\cite{gw-astro-PE-lalinference-v1,PEPaper,NRPaper} and references therein).
In this expression $x$ is shorthand for the 15 parameters needed to fully specify a quasicircular BH-BH binary in space and
time, relative to our instrument, and $d$ denotes all the gravitational wave data from all of LIGO's instruments.  

Using Bayes' theorem, the posterior distribution is proportional to the product of this likelihood and our prior
assumptions $p(x)$ about the probability of finding a black hole merger with different masses, spins, and orientations
somewhere in the universe [$p(x|d)\propto p(d|x)p(x)$].    LIGO adopted a fiducial prior $p_{\rm ref}(x)$, uniform in
orientation, in comoving volume, in mass,  in spin direction (on the sphere), and, importantly for us, in spin magnitude~\cite{gw-astro-PE-lalinference-v1,PEPaper}.   

Using standard Bayesian tools~\cite{PEPaper,gw-astro-PE-lalinference-v1}, one can produce a sequence of independent, identically distributed samples
$x_{n,s}$ ($s=1,2,\ldots,S$) from the posterior distribution $p(x|d)$ for each event $n$; that is, each $x_{n,s}$ is
drawn from a distribution proportional to $p(d_n|x_n)p_{\rm ref}(x_n)$.   Typical calculations of this type  provide  $\lesssim 10^4$
samples~\cite{PEPaper,gw-astro-PE-lalinference-v1}. %
Using classical distribution and density estimation techniques, the samples let us infer the binary parameter
distributions.  For example, if $X$ is some scalar quantity (e.g., the chirp mass) derived from $x$, then  the
cumulative distribution $P(<X)$ can be estimated by $\sum_k \Theta(X-X_k)/N$, where $\Theta$ denotes the Heavyside step function.

\begin{figure*}
\includegraphics[width=\columnwidth]{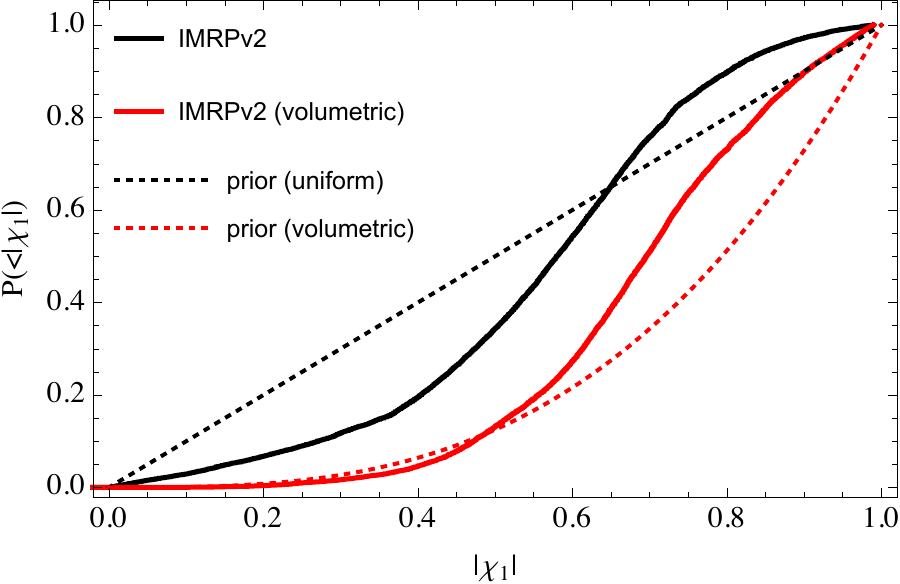}
\includegraphics[width=\columnwidth]{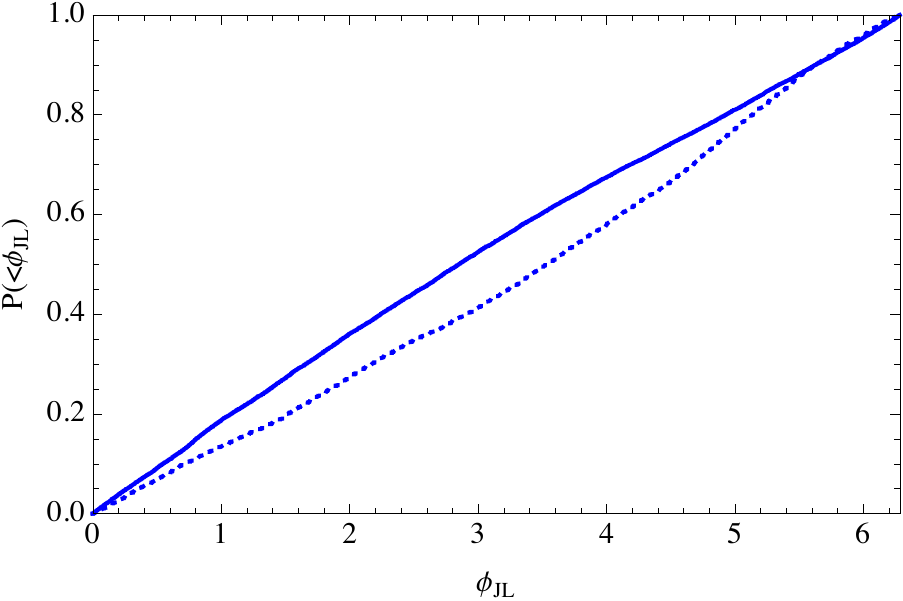}
\caption{\label{fig:PE:Priors:GW151226}\textbf{Priors and the relative significance of large spins} %
\emph{Left panel}:
 For a synthetic GW151226-like event, the inferred cumulative posterior distribution for $|\chi_1|$ using a prior $P(<|\chi_i|)=\chi_i$ (black)
and $P(<|\chi_i|)=\chi_i^3$ (red), for $i=1,2$.  For comparison, the two priors are indicated with dotted black and red lines.  The
posterior probability that this synthetic event has two significant, precessing spins depends  on the prior.
\emph{Right panel}: Inferred cumulative posterior distribution for $\phi_{JL}$, the polar angle of $\mathbf{L}$ relative
to $\mathbf{J}$, for the
volumetric prior $P(<|\chi_i|)=\chi_i^3$.  The solid blue line shows the results of repeating a full PE calculation,
including the modified prior.  The dotted blue line shows the estimated distribution calculated by weighting the
posterior samples.
This synthetic event was generated with parameters similar to GW151226
    and analyzed with a PSD appropriate to GW150914, generated in the manner of~\cite{Abbott:2016apu}. 
}
\end{figure*}

As is immediately apparent from Bayes' theorem and has long been understood, the choice of prior $p_{\rm ref}$ directly influences the
posterior, particularly for parameters not well constrained by the data (e.g., due to weak depedence or strong
degeneracies).   As a concrete example, in the left panel of Figure \ref{fig:PE:Priors:GW151226} we show 
the cumulative distribution of $\chi_{1,z}$ for  a synthetic
  source similar to GW151226.  
The black curve corresponds to results evaluated using the fiducial prior, where $\chi_1,\chi_2$ are
distributed independently and 
uniformly.  The red curve is computed by drawing $\chi_1$ from the cumulative distribution $P(<\chi_1) = \chi_1^3$ and
similarly from $\chi_2$, which we henceforth denote the volumetric (spin) prior.  
In the context of systematic errors and astrophysical measurements of BH-BH binaries, the choice of prior is important.
Within the context of a specific astrophysical scenario or question of interest, a prior favoring large spins (or
significant precession) can be appropriate.  As we show below, these changes in prior can significantly increase the
posterior probability of the region where model disagreement is substantial (e.g., large transverse spins, high mass
ratio, and long signals).

When assessing the impact of modified priors, we exercise an abundance of caution and replicate the Bayesian
inference calculations in full.  In principle, with sufficiently many samples, we could estimate the posterior
distribution for any prior $p(x)$  by using weighted samples.  For example, we could estimate $P(<X)$ according to the
modified prior $p(x)$ via the weighted empirical cumulative distribution $\hat{P}(<X) = \sum_k \Theta(X-X_k)
p(x_k)/(Np_{\rm ref}(x_k))$.  The approach of reweighted posterior samples is widely proposed in hierarchical model
selection~\cite{2010ApJ...725.2166H,2017arXiv170601385F}.
 In practice, however, this method is reliable if and only if $x_k$ cover the parameter
space completely and sufficiently densely. In our specific circumstances, the fiducial prior $p_{\rm ref}(x)$ associates substantial prior weight near $\chi_1,\chi_2\simeq 0$
and little probability to configurations with two large spins.  As a result, rescaling from the fiducial to the
volumetric prior can introduce  biases into astrophysical conclusions.  As a concrete example, the right panel of Figure 
\ref{fig:PE:Priors:GW151226} shows the cumulative distribution of $\phi_{JL}$, the polar
angle of $\mathbf{L}$ relative to $\mathbf{J}$. %
The solid line shows the result of a full calculation with the volumetric prior.  The dotted line shows the result derived using
reweighted posterior samples, starting from the fiducial uniform-magnitude prior.   While the two distributions are
approximately consistent in extent, the two disagree in details.   
If used uncritically in (hierarchical) model selection, reweighted posterior samples could lead to biased conclusions
about model inference, and (in the context of our study) to biased conclusions about the relative impact of model-model systematics. 
Of course,  a careful treatment of reweighted posterior systematics would identify this potential problem, and the need
for more samples to insure a reliable answer in any reweighted application (i.e., the expected variance of the Monte Carlo integral estimate for $\hat{P}$ is large, because
$p/p_{\rm ref}$ is often large). %

\subsection{Model-model comparisons}
To quantify the difference between two predicted gravitational waves from the same binary with the same spacetime
coordinates and location, we use a standard
data-analysis-motivated figure of merit: the mismatch.
Like other figures of merit, the mismatch is calculated using an inner product %
between two (generally complex-valued) timeseries $a(t),b(t)$:
\begin{equation}
    \qmstateproduct{a}{b} = 2 \int_{|f| \geq f_{min}} \frac{\tilde{a}^*(f) \;  \tilde{b}(f)}{S_n(|f|)} df \, ,
\end{equation}
where $S_n(|f|)$ is the noise power spectral density of a fiducial detector,  $f_{min}$ is a chosen lower frequency
cut-off (typically a few tens of Hz), and the integral includes both positive and negative frequencies.
Usually these comparisons also involve parameterized signals
$a(\lambda,\theta)$ and $b(\lambda',\theta')$, with  maximization of the (normalized) inner product $P$ between $a,b$ over some set of parameters $\theta$:
\begin{eqnarray}
P( a,b | \Theta) = \text{max}_\theta \frac{\text{Re}\qmstateproduct{a(\theta)}{b(\theta')}}{\sqrt{\qmstateproduct{a(\theta)}{a(\theta)}\qmstateproduct{b(\theta')}{b(\theta')}}}
\end{eqnarray}
where $\Theta$ denotes the names of the parameters in $\theta$ over which we maximize.  Maximization is asymmetric; we change the parameters of only one of the two signals, effectively considering the other as ``the source''.  
When the signals $a,b$ are real-valued single-detector response functions and when $\Theta$ is time and binary orbital phase, this expression is known as the
\emph{match}.  When the two signals are real-valued single-detector response functions and when $\Theta$ includes all binary parameters, this expression is known
as the \emph{fitting factor}.

In our comparisons, we fix  one of the two timeseries $a$ generated by model $A$, as if it was  some known detector response (e.g., from
   another model's prediction). The other timeseries is a  predicted single-detector
   response $b= \text{Re} F^* h$ where $F$ is a complex-valued antenna response function and $h$ is the gravitational
   wave strain.  Ideally, we should evaluate $b$ using model $B$ and  precisely the same intrinsic and extrinsic
   parameters, calculating the \emph{faithfulness}~\cite{1998PhRvD..57..885D}. 
The
precessing models considered in this work have different time and phase conventions.  In order to specify the
astrophysically equivalent binary to some configuration as evolved by \SEOB, we need to adopt a different event time
$t$; orbital phase $\phi_{\rm orb}$; and precession phase $\phi_{JL}$. 
Reconciling the phase conventions adopted by these models is far beyond the scope of this work.  However, we can find
the most optimistic  possible answer by maximizing overlap over $t$, $\phi_{\rm orb}$, and $\phi_{JL}$, using
differential evolution to evolve towards the best-fitting signal.  In other words, we use a figure of merit
\begin{eqnarray}
    \label{eq:fom}
    P(a,b | t, \phi_{\rm orb}, \phi_{JL})
\end{eqnarray}
Because our gravitational wave signals include higher modes, we take care \emph{not} to use the conventional Fourier
transform trick when maximizing in time and orbital phase. 

\subsection{Comparison on posterior distributions}
To  investigate systematic errors in observationally relevant regions of parameter space,%
we perform model-model comparisons using samples drawn from the posterior parameter distributions for several of LIGO's
detections to date, as well as for synthetic sources.   Unless otherwise noted, these comparisons are performed on the
expected Hanford detector response.

\begin{figure*}
    \includegraphics[width=\columnwidth]{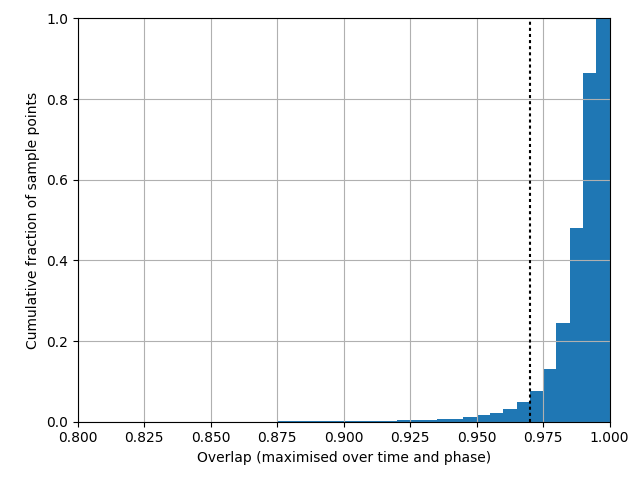}
    \includegraphics[width=\columnwidth]{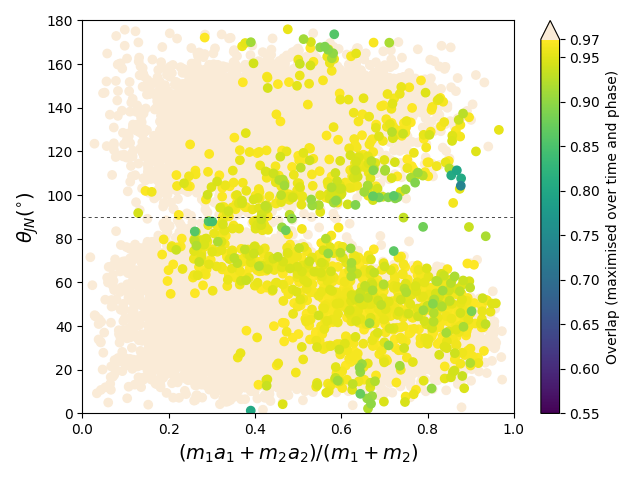}
    \caption{
        \label{fig:170104:NoVolumetric}
        \textbf{Model-model comparison on our synthetic GW170104-like event}:
        Using posterior samples from our synthetic GW170104-like event, we calculate model-model overlaps between \IMRP and \SEOB  waveforms, maximized over time, coalescence phase, and precession phase.  
        This analysis adopts the fiducial (uniform) prior on spin magnitude. 
        In the left panel is a cumulative histogram of the maximized overlaps.
        In the right panel the posterior samples are plotted in terms of $\theta_{JN}$, the inclination of the observer relative to the total angular momentum, and a measure of the net binary BH spin.
        The color scale indicates the maximized overlap, with the lowest values occurring for large binary spins and preferentially near the orbital plane.
        The noise curve used for these calculations was the same as used in Figure \ref{fig:PE:Priors:GW151226}.
    }
\end{figure*}

Figure \ref{fig:170104:NoVolumetric} illustrates our comparisons for our synthetic GW170104-like event, using the fiducial spin prior.
For the short waveforms needed to explain this signal, disagreement between the models is primarily associated with
higher levels of precession, viewed in an orientation near the orbital plane where the effects of precession dominate.

\begin{figure*}
    \includegraphics[width=\columnwidth]{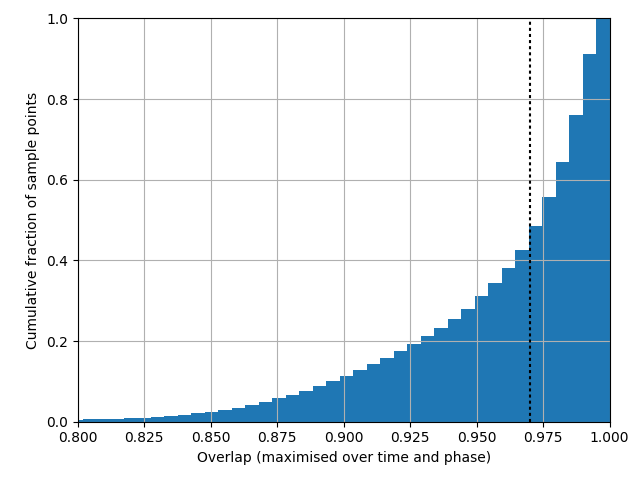}
    \includegraphics[width=\columnwidth]{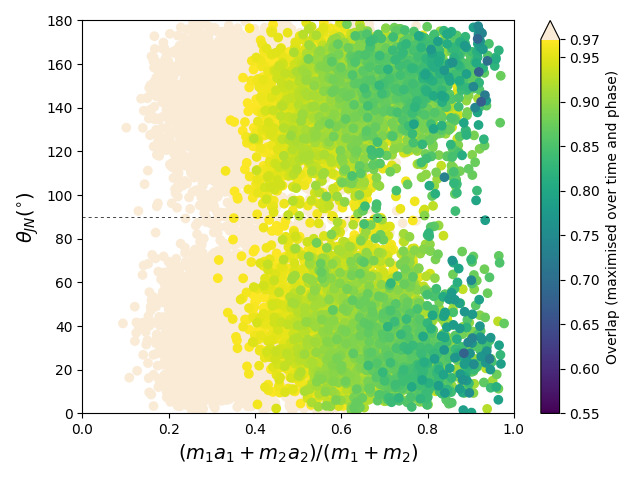}
    \caption{\label{fig:151226}
        \textbf{Model-model comparison on GW151226}: As Figure \ref{fig:170104:NoVolumetric} but for GW151226.  The intrinsic and
        extrinsic parameters used in this comparison are from LIGO's  O1 posterior distribution.
        Frequent and significant  disagreement is apparent.
        \IMRP produces waveforms that are somewhat longer than \SEOB for these modest masses, leading to dephasing due to a slight difference in the rate of frequency evolution integrating over such long waveforms.
        This effect correlates strongly with the binary spin.
        The noise curve used for these calculations was calculated from data near the time to GW 151226.
    }
\end{figure*}

Since lower mass systems take longer to evolve from some lower frequency to the merger, waveforms drawn from the posterior of GW151226 are significantly longer in duration.
The two waveform models have significantly greater opportunity to dephase, leading to lower overlaps.    
Figure \ref{fig:151226} shows the distribution of these mismatches.
The disagreement is significant over a larger portion of the parameter space than for our GW170104-like event, and is less strongly correlated with the orbital inclination $\theta_{JN}$.
It is clear that the spins play a leading role in producing these differences.

\begin{figure*}
    \includegraphics[width=\columnwidth]{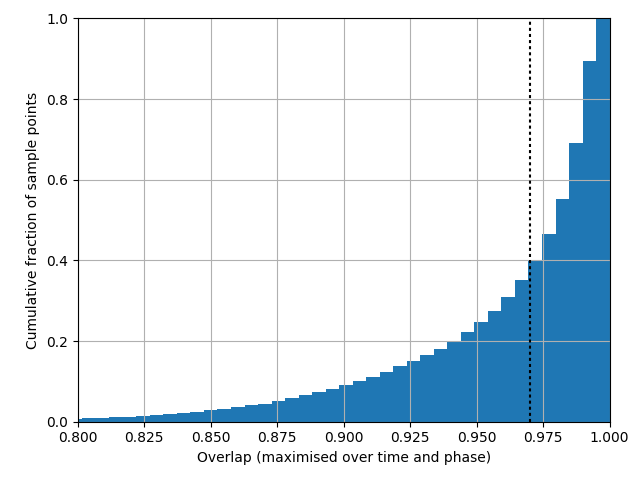}
    \includegraphics[width=\columnwidth]{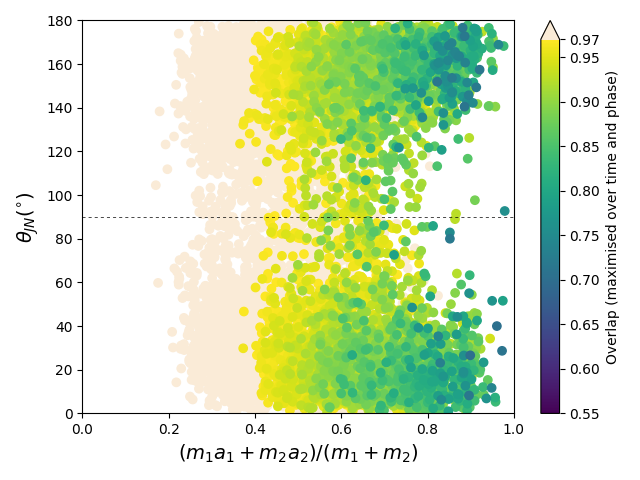}
    \includegraphics[width=\columnwidth]{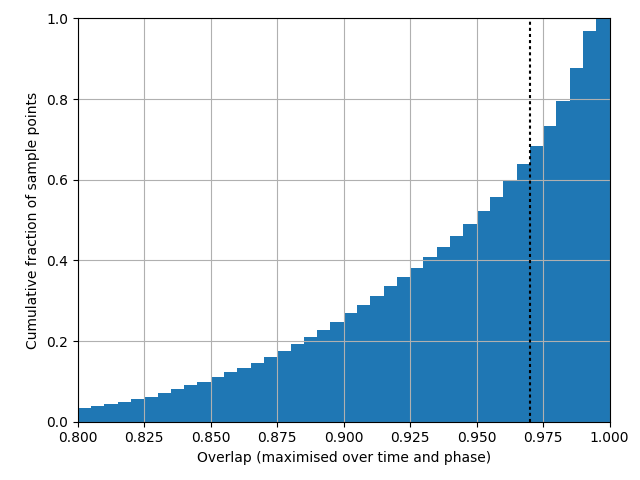}
    \includegraphics[width=\columnwidth]{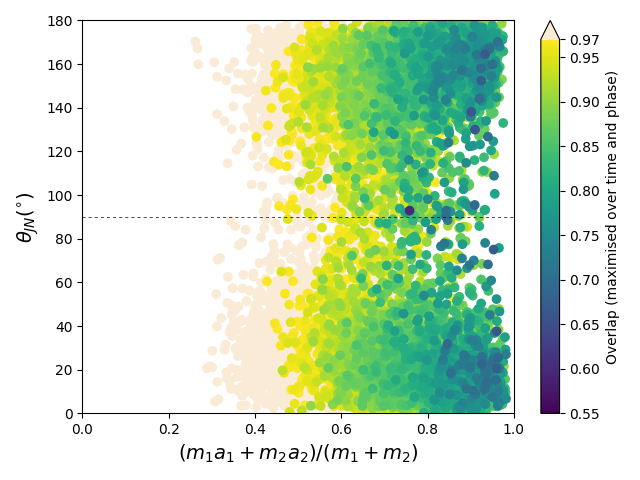}
    \caption{\label{fig:151226:Synth}
        \textbf{Model-model comparison on a synthetic GW151226-like event}: As Figures \ref{fig:170104:NoVolumetric} and
        \ref{fig:151226} but for a
        synthetic GW151226-like event.  As in Figure \ref{fig:170104:NoVolumetric}, the top and bottom panels show the results assuming a
        uniform and volumetric spin prior, respectively.  Adopting a volumetric spin prior noticably increases the
        posterior support for large spins and hence the fraction of the posterior associated with parameters where the two models disagree significantly. 
    }
\end{figure*}

In Figure \ref{fig:151226:Synth} we show two results for a synthetic GW 151226-like event, one that adopts a uniform spin prior, and the other the volumetric spin prior.
As in Figure \ref{fig:151226} we see that more moderate-to-highly spinning systems in the posteriors show a greater degree of diagreement between the models.
The volumetric spin prior increases the support for large spins, and so increases the proportion of the posterior where model disagreement is significant.

\section{Examples of biased inference of BH parameters}
\label{sec:PE:Examples}

To illustrate  the discrepancies in inferred parameters which such disagreements can cause, we select points with significant differences and generate the associated waveforms with one model, before running the full parameter estimation analysis on these waveforms using the other model.
We do not add any simulated instrumental noise to the model signal in this process.

\begin{figure*}
\includegraphics[width=0.32\textwidth]{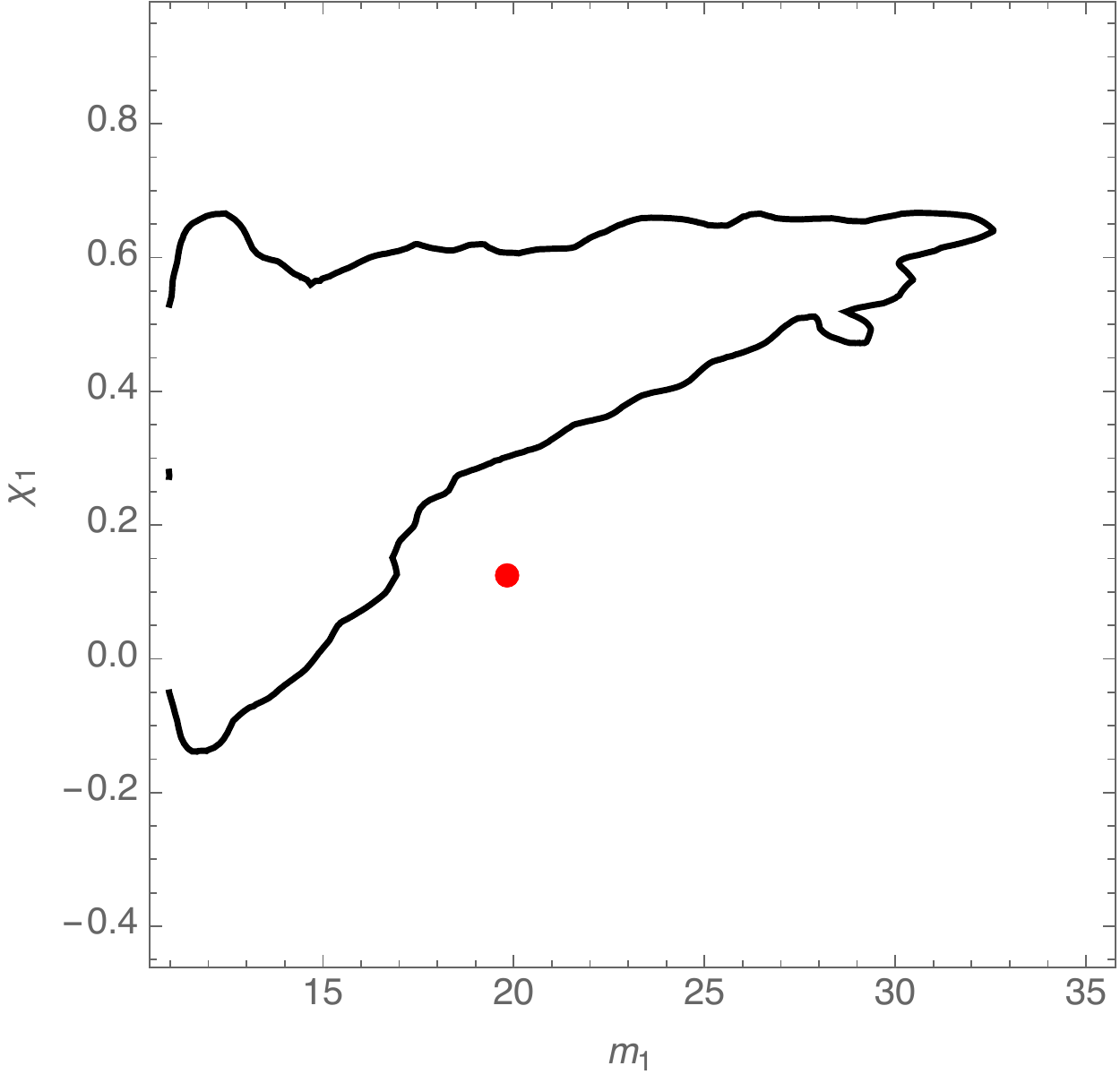}
\includegraphics[width=0.32\textwidth]{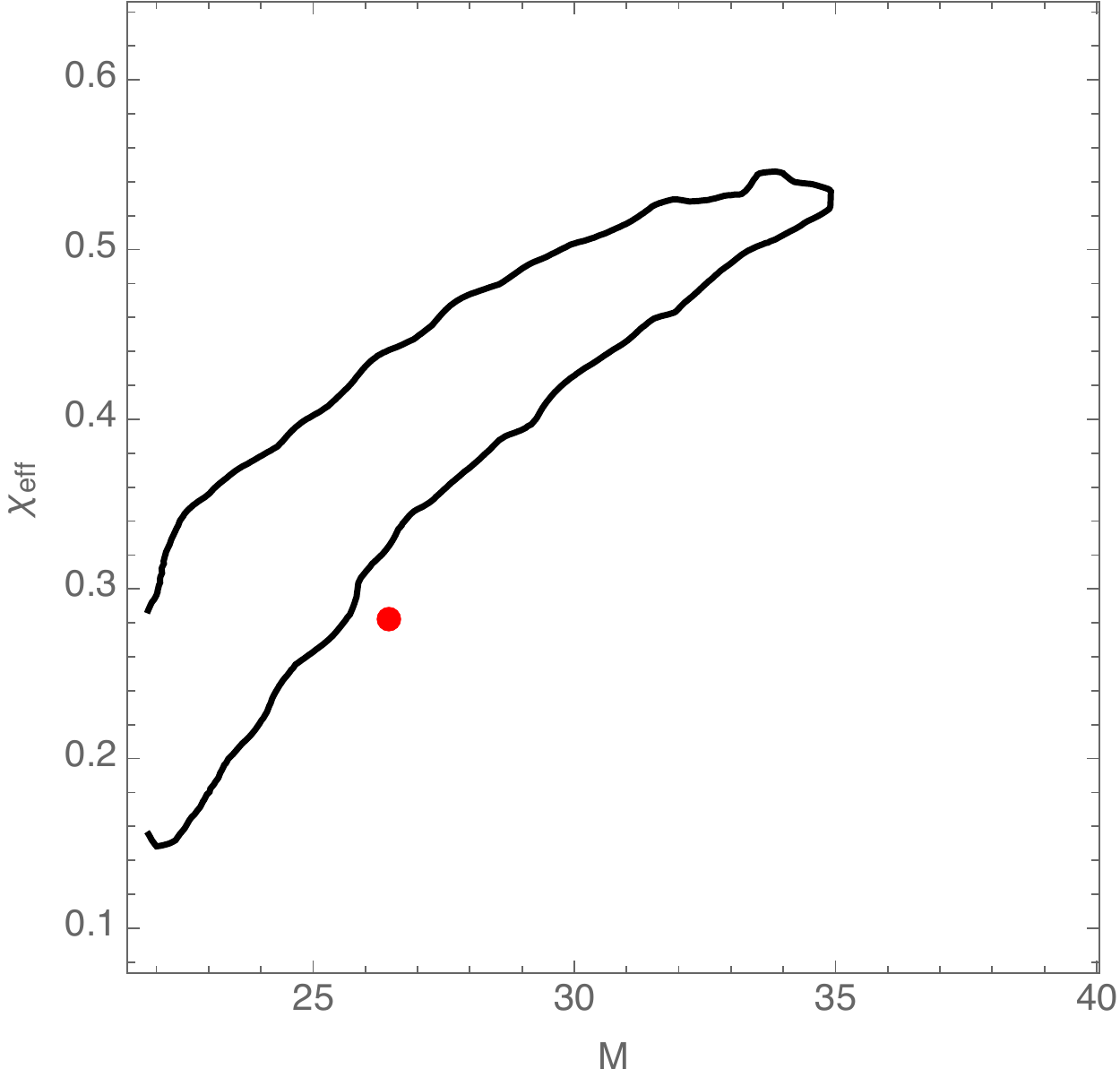}
\includegraphics[width=0.32\textwidth]{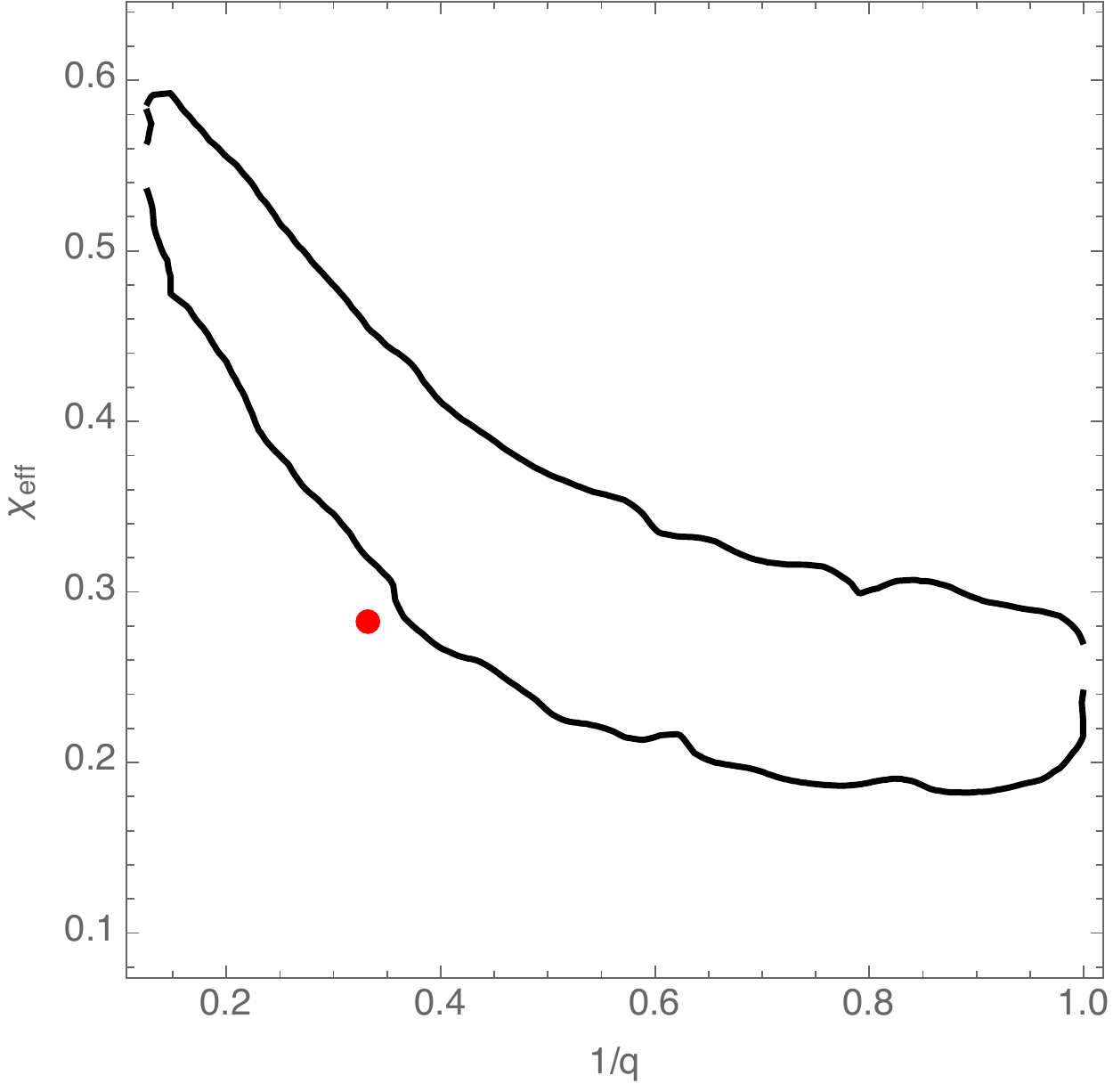}
\caption{\label{fig:BiasedPE}\textbf{Biased parameter recovery with \IMRP I: \SEOB source}: Red dot shows the parameters of a synthetic coalescing binary,
  whose radiation is modeled with \SEOB.  Binary parameters are drawn from the posterior distribution of
  GW151226.  The inclination of the orbital angular momentum relative to our line of sight is $\theta_{JN}=2.48$.   No synthetic
  noise is added to the signal.  For this source, the match between the detector response predicted using \IMRP and
  \SEOB is 0.817 in Hanford, after maximizing in $t,\phi_{orb},\phi_{JL}$.
  Black curve shows the 90\% confidence interval derived from a detailed parameter inference calculation using the
  \IMRP approximation. 
 Calculations are performed using a network of detectors whose noise power spectra are identical to the estimates
    derived for GW150914~\cite{LIGO-O1-BBH,Abbott:2016apu}, using frequencies above $20\unit{Hz}$. 
}
\end{figure*}

\begin{figure*}
\includegraphics[width=0.32\textwidth]{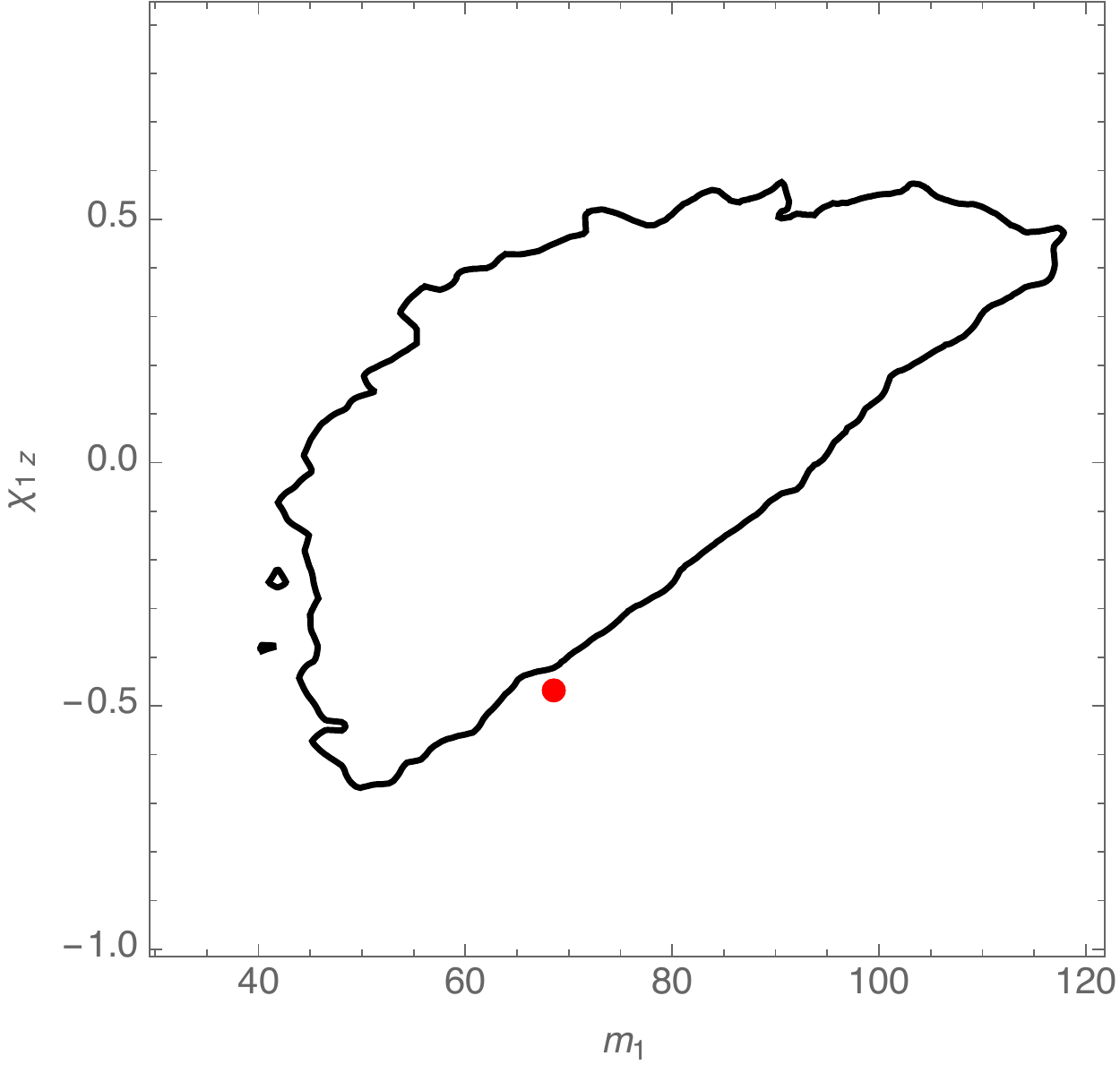}
\includegraphics[width=0.32\textwidth]{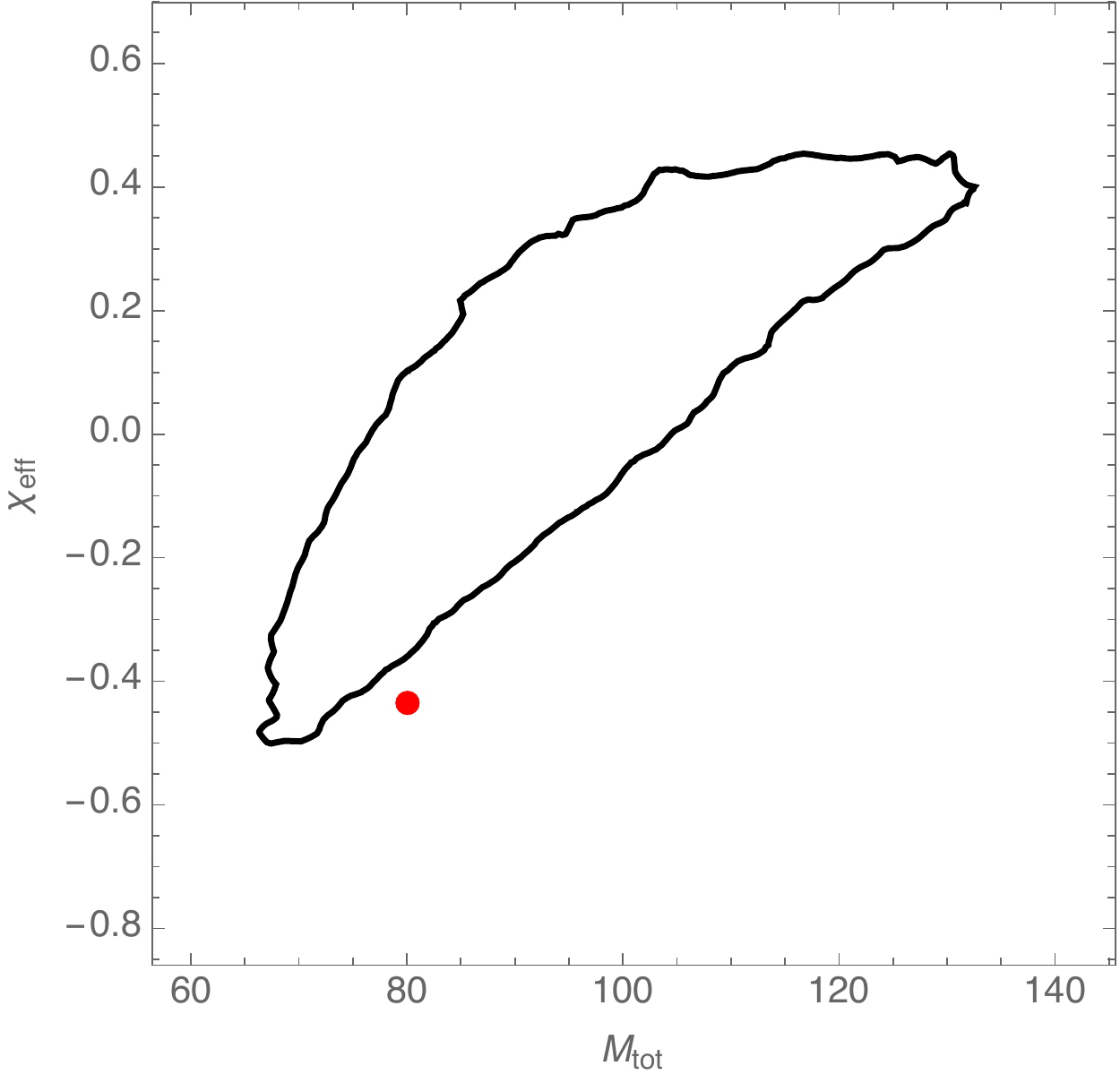}
\includegraphics[width=0.32\textwidth]{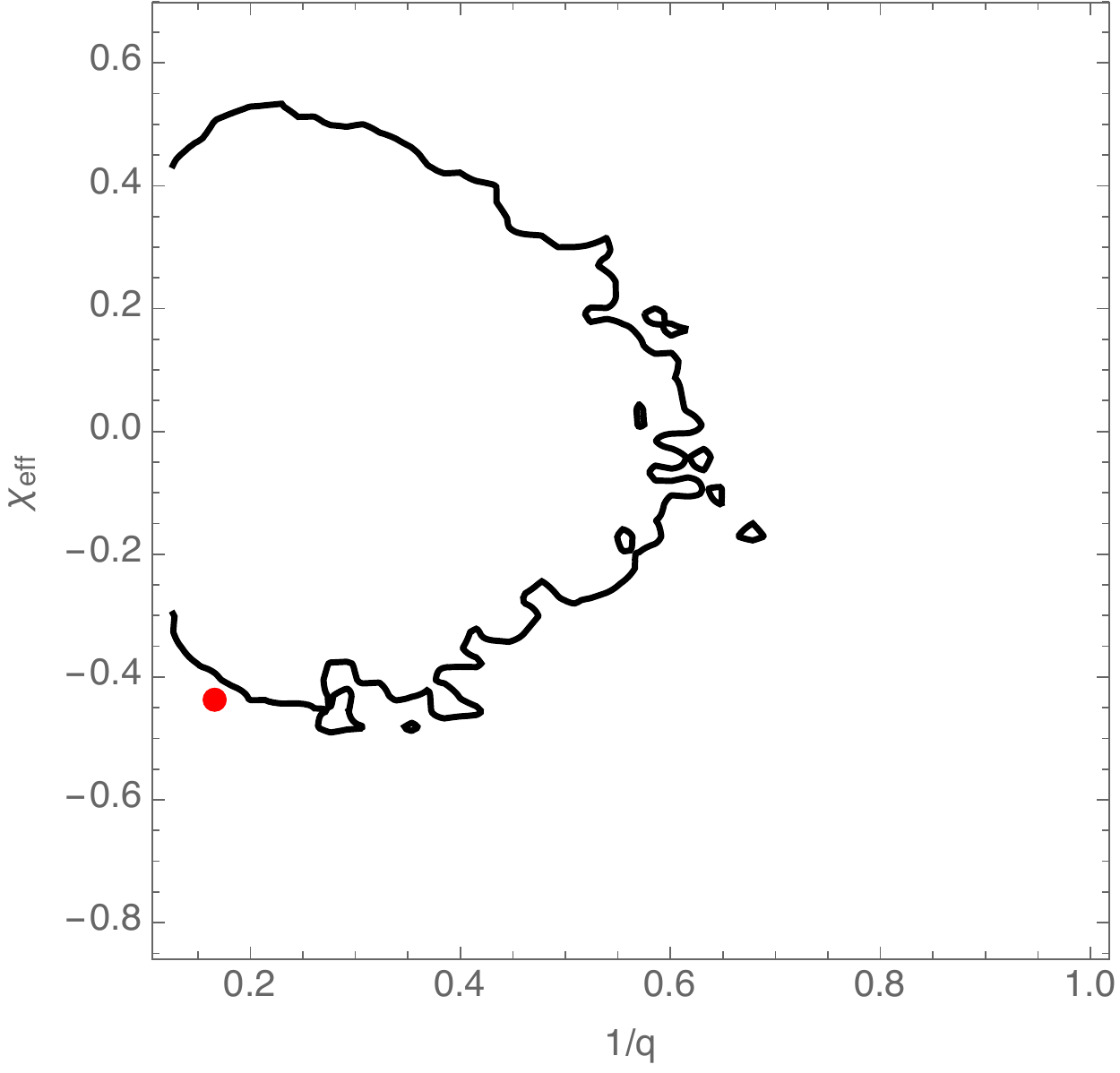}
\caption{\label{fig:BiasedPE:NR}\textbf{Biased parameter recovery with \IMRP II: NR source}: Red dot shows the parameters of a synthetic coalescing binary,
  whose radiation is modeled with a numerical relativity simulation SXS BBH:0165 with network SNR of 12.    All simulated modes $\ell \le 8$
  are included in our synthetic signal.    The detector response is calculated assuming a signal an angle  $\theta_{JN}=\pi/ 4$,
    at a distance so the network SNR is ${\sim}10$.
    No synthetic noise is added. 
    The black curves show the 90\% confidence interval derived from a detailed parameter inference calculation using the \IMRP approximation. 
    Calculations are performed using a network of detectors whose noise power spectra are identical to the estimates
    derived for GW150914~\cite{Abbott:2016apu}.  Because the $(2,2)$ mode of this source starts at $27\unit{Hz}$, we only use
  frequencies greater than $30\unit{Hz}$ in our analysis
}
\end{figure*}

In these  demonstrations of the practical differences between models from each other and from numerical relativity, we use
the same parameter estimation techniques and models applied by LIGO to infer the parameters of the first two observed binary black
holes~\cite{2015PhRvD..91d2003V,2016PhRvL.116x1102A,LIGO-O1-BBH}.  
Figures \ref{fig:BiasedPE} and \ref{fig:BiasedPE:NR} show  concrete examples of biased parameter inference.   In Figure
\ref{fig:BiasedPE}, the red dot shows
the parameters of a synthetic signal, generated with \SEOB using intrinsic and extrinsic parameters drawn from the posterior distribution for
GW151226.  The binary parameters chosen correspond to a configuration where the models disagree (i.e., low overlap); see
Figure  \ref{fig:151226}.  Our synthetic data contains only the expected detector response, with no noise (the ``zero
noise'' realization).  
The black curves show the 90\% posterior confidence intervals, derived using the \IMRP parameter inference engine.
In Figure \ref{fig:BiasedPE:NR}, we generate a synthetic source signal from a numerical relativity simulation
produced by the SXS collaboration~\cite{2013PhRvL.111x1104M}, using an extension to LIGO's infrastructure to designed for this purpose~\cite{2017arXiv170301076S,2016arXiv161107529G}. 
This figure demonstrates by concrete example that the two models' disagreement can propagate into biased inference about astrophysically
important binary parameters, even now in a regime of low signal amplitude and large statistical error.

\section{Discussion}
\label{sec:Discussion}

\subsection{Mismatch does not imply bias: Examples with high mass ratio and zero spin}
\begin{figure}
    \includegraphics[width=\columnwidth]{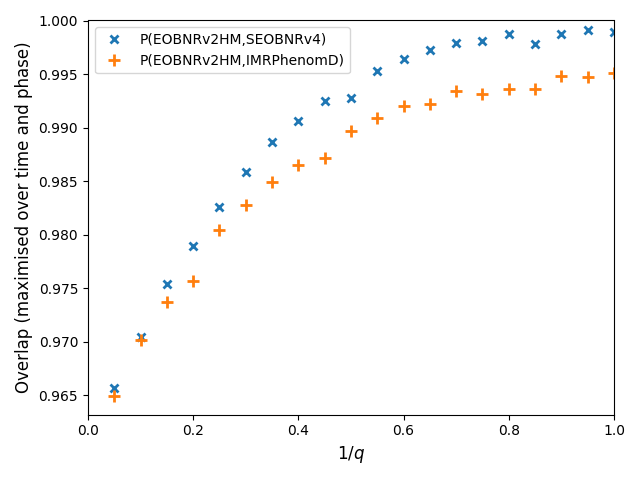}
    \caption{\label{fig:overlap_v_q}
        \textbf{The effect on overlap due to neglecting higher modes}:
        Here we generate a series of non-spinning waveforms with $M=80 M_\odot$ and $\theta_{JN}=\pi/ 4$ using an EOB model that includes higher modes, \textsc{EOBNRv2HM}, then use the same parameters to generate waveforms with two
        models that do not include these higher modes, one EOB and one phenomenological -- \textsc{SEOBNRv4} and \textsc{IMRPhenomD}.
        Again we calculate the overlap maximized over $\phi_{\mathrm{orb}}$, $\phi_{JL}$, and $t$.  
      As higher modes are most important for heavier and unequal mass binaries, these large mismatches may be
      responsible for disagreements seen in Figure \ref{fig:BiasedPE:NR}.  Conversely, higher  modes are not significant for and not included in 
      models compared Figures \ref{fig:151226} and \ref{fig:BiasedPE}, so are unlikely to be responsible for the large
      discrepancies seen there.
    }
\end{figure}

Due to their neglect of higher-order modes, the two models disagree significantly with numerical relativity at high mass
ratio, even in the absence of spin.  Several previous studies have demonstrated these modes have a significant impact on
the match~\cite{2014PhRvD..90l4004V,2016PhRvD..93h4019C,2016arXiv161205608V,2017PhRvD..95j4038C}. %
As a concrete example, Figure \ref{fig:overlap_v_q} illustrates the mismatch introduced due to the neglect of higher order modes for non-spinning systems of varying mass ratios, with total masses $M=80 M_\odot$ and inclinations $\theta_{JN}=\pi/ 4$.

\begin{figure*}
\includegraphics[width=0.32\textwidth]{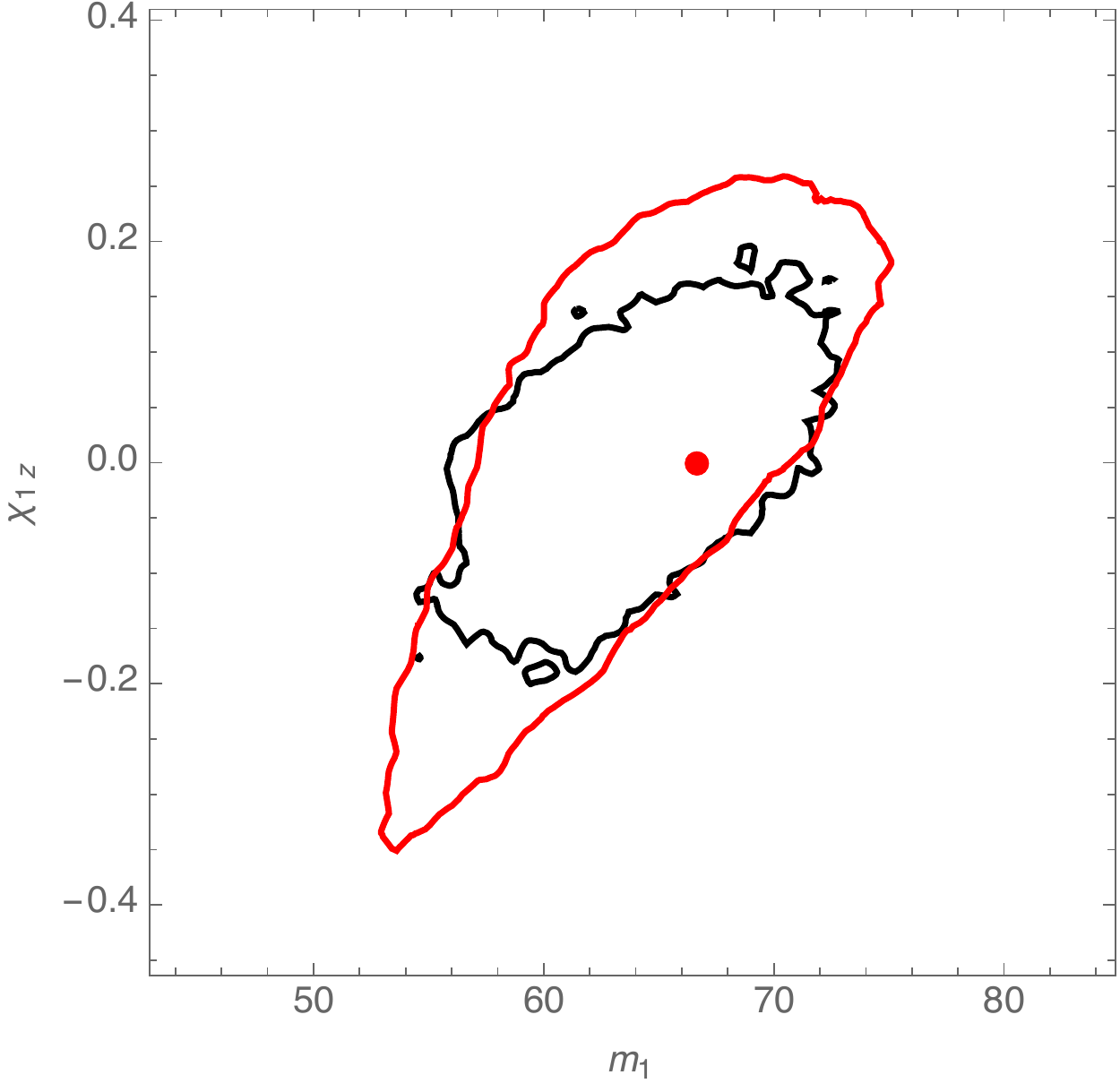}
\includegraphics[width=0.32\textwidth]{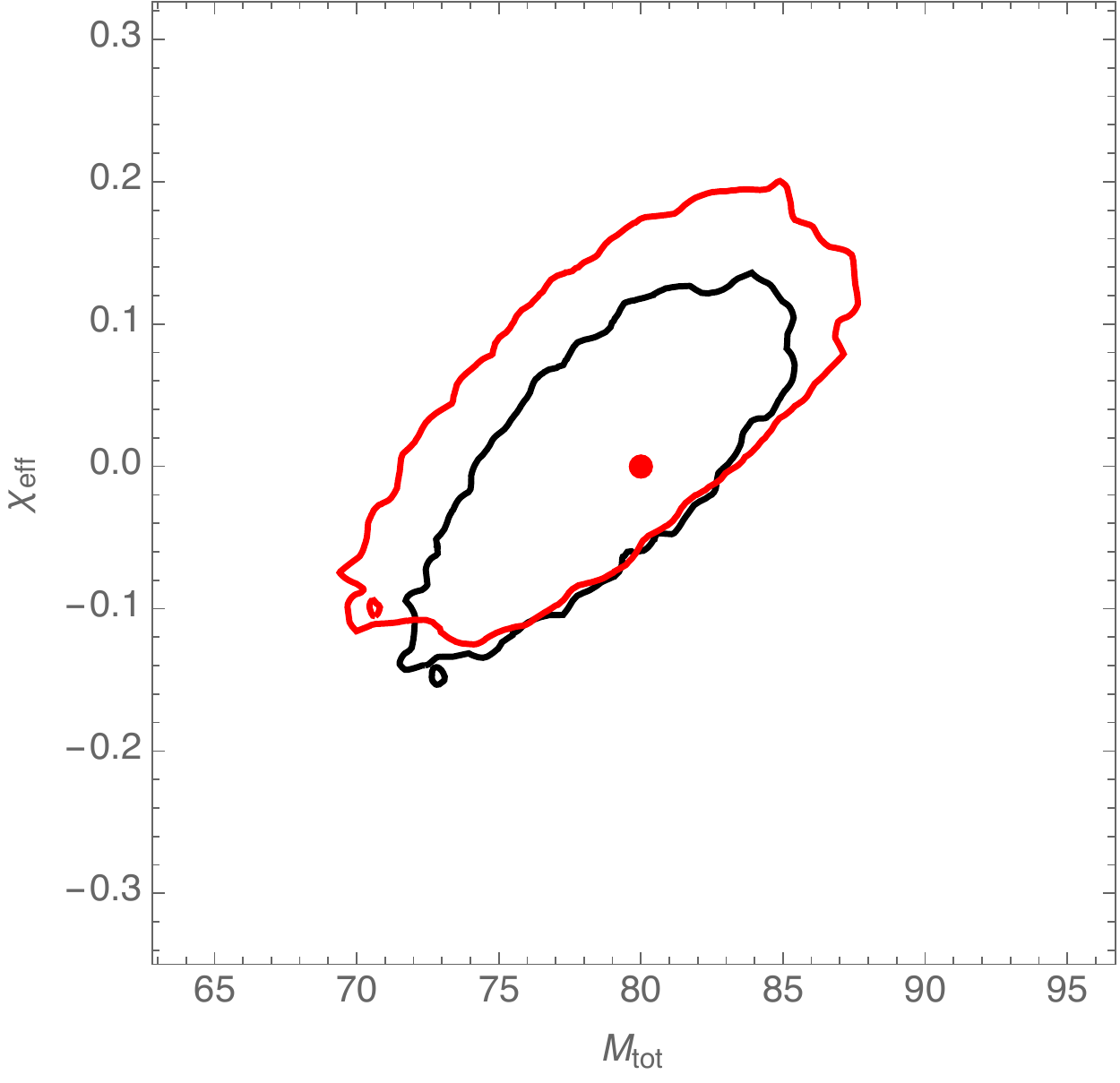}
\includegraphics[width=0.32\textwidth]{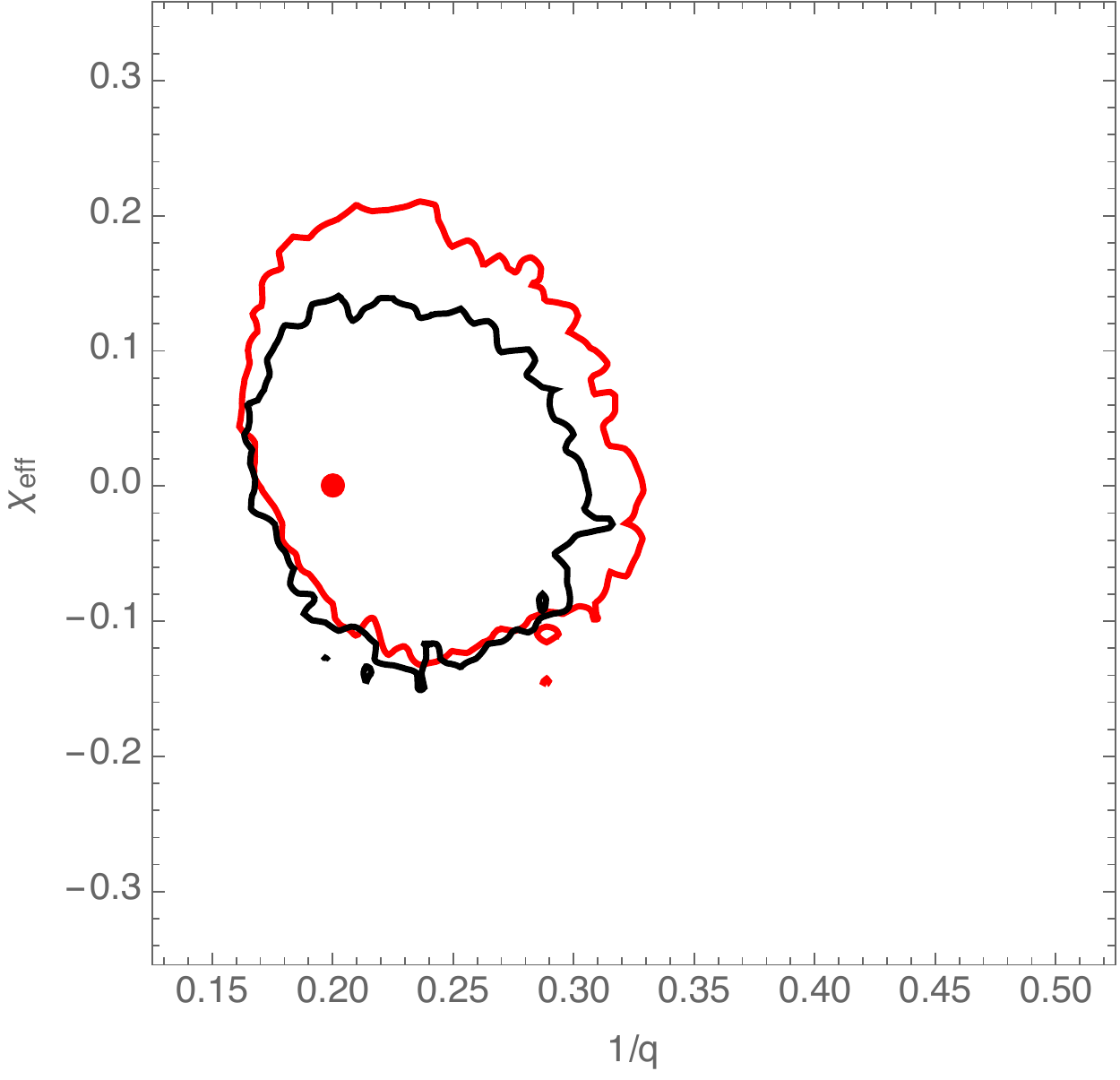}
\caption{\label{fig:NoBiasPE:HigherModeExample}\textbf{Omitting higher modes: Unbiased parameter inference, despite a high mismatch}: Red dot shows
  the parameters of a synthetic nonprecessing  binary,
  whose radiation is modeled with a numerical relativity simulation SXS BBH:0112.    All simulated modes $\ell \le 8$
  are included in our synthetic signal.    The detector response is calculated assuming a source with total mass $80 M_\odot$ oriented at angle  $\theta_{JN}=\pi/ 4$,
  at a distance so the network SNR is 20. No synthetic   noise is added.  For this source, the best match with the \IMRP
  and \SEOB approximations is $\simeq 0.96$.
  The black and red curves  shows the 90\% confidence interval derived from a detailed parameter inference calculation using the
    \IMRP and \textsc{SEOBNRv4} approximations, respectively. 
 Calculations are performed using a network of detectors whose noise power spectra are identical to the estimates
    derived for GW150914~\cite{Abbott:2016apu}.
}
\end{figure*}

A large mismatch, however, does not imply a large \emph{bias}. 
 For example, in the limit of a long signal, the
different harmonics have distinct time-frequency trajectories and transfer information with minimal
cross-contamination~\cite{gw-astro-SpinAlignedLundgren-FragmentA-Theory,gwastro-SpinTaylorF2-2013,gwastro-mergers-Prakash-PrecessingFisherMatrix-2014}.   As a result, an analysis using only \emph{one} mode will find similar best-fitting parameters, but
with a wider posterior than if all available information was used.   
At higher mass and near the end of the merger, however, multiple modes are both significant and, due to their brevity,
harder to distinguish.   
Using a simple matched-based analysis applied to hybridized nonprecessing multimodal NR simulations,~\cite{2016arXiv161205608V} argued that for moderate-mass binaries, inferences  based on the leading-order quadrupolar model alone would not be
significantly biased, compared to the (large) statistical error expected at modest SNR; see the right panel of their
Figure 1. 
For nonprecessing zero-spin binaries, we confirm by example that inferences about the binary are not biased.
As an example, Figure   \ref{fig:NoBiasPE:HigherModeExample}  
shows the posterior distributions inferred using two EOB models, one including higher-order modes (\textsc{EOBNRv2HM}), and the other omitting them (\textsc{SEOBNRv4}).
The synthesised signal is a nonprecessing binary with $q=5$ and $M=80 M_\odot$, generated via numerical relativity (i.e., a signal including higher order modes).
Due to model limitations,
these inferences are performed assuming both black holes have zero spin.   This figure shows that both sets of parameter
inferences are consistent with the true binary parameters used, and that inferences constructed with higher modes (via
\textsc{EOBNRv2HM}) are both sharper 
and less biased than inferences that omit higher modes (via \textsc{SEOBNRv4}).

A large mismatch does imply, however, that the analysis is not using all available information.   
For example, searches for gravitational waves which neglect  higher modes cannot fully capture all available signal
power and a priori are somewhat less sensitive~\cite{2010PhRvD..82j4006O,2016arXiv161205608V,2016PhRvD..93h4019C,2017PhRvD..95j4038C}; but
cf.~\cite{2014PhRvD..89j2003C}.  
Parameter inference   calculations that use higher modes are well-known to be more discriminating about binary
parameters~\cite{2007CQGra..24..155V,2006PhRvD..74l2001L,2009PhRvD..80f4027K,2008PhRvD..78f4005P,2011PhRvD..84b2002L,2011PhRvD..84b2002L,2014PhRvD..89f4048O,2015PhRvD..92b2002G,2017CQGra..34n4002O}.  Even for short signals associated with heavy binary black holes, analyses with higher modes can draw
 tigher inferences about binary parameters~\cite{2015PhRvD..92b2002G,2017CQGra..34n4002O,NRPaper}, depending on the
 source; see, e.g., Figure  \ref{fig:NoBiasPE:HigherModeExample}.

\subsection{Marginal distributions, degeneracy, and biases}

Fortunately or not, nature and LIGO's instruments have conspired to produce short GW signals with modest amplitude to date.
As illustrated by LIGO's results~\cite{LIGO-O1-BBH,2017PhRvL.118v1101A} and our Figure \ref{fig:BiasedPE}, when using current methods (e.g., \IMRP and \SEOB), the
inferred posterior distributions  for most parameters are quite broad, dominated by substantial statistical error.    
Inferences about individual parameters are also protected by strong degeneracies in these  approximate models (e.g., in
the neglect of higher-order modes) and in the physics of binary mergers (e.g., in the dependence of merger trajectories
on net aligned spin).  For example, the rightmost panel of Figure \ref{fig:BiasedPE} shows the posterior distribution in
$q$ and $\chi_{\rm eff}$; the joint posterior is tightly correlated (and strongly biased), but the individual marginal distributions for $q$
and $\chi_{\rm eff}$  is broad, and contains the true parameters.    

\begin{figure}
\includegraphics[width=\columnwidth]{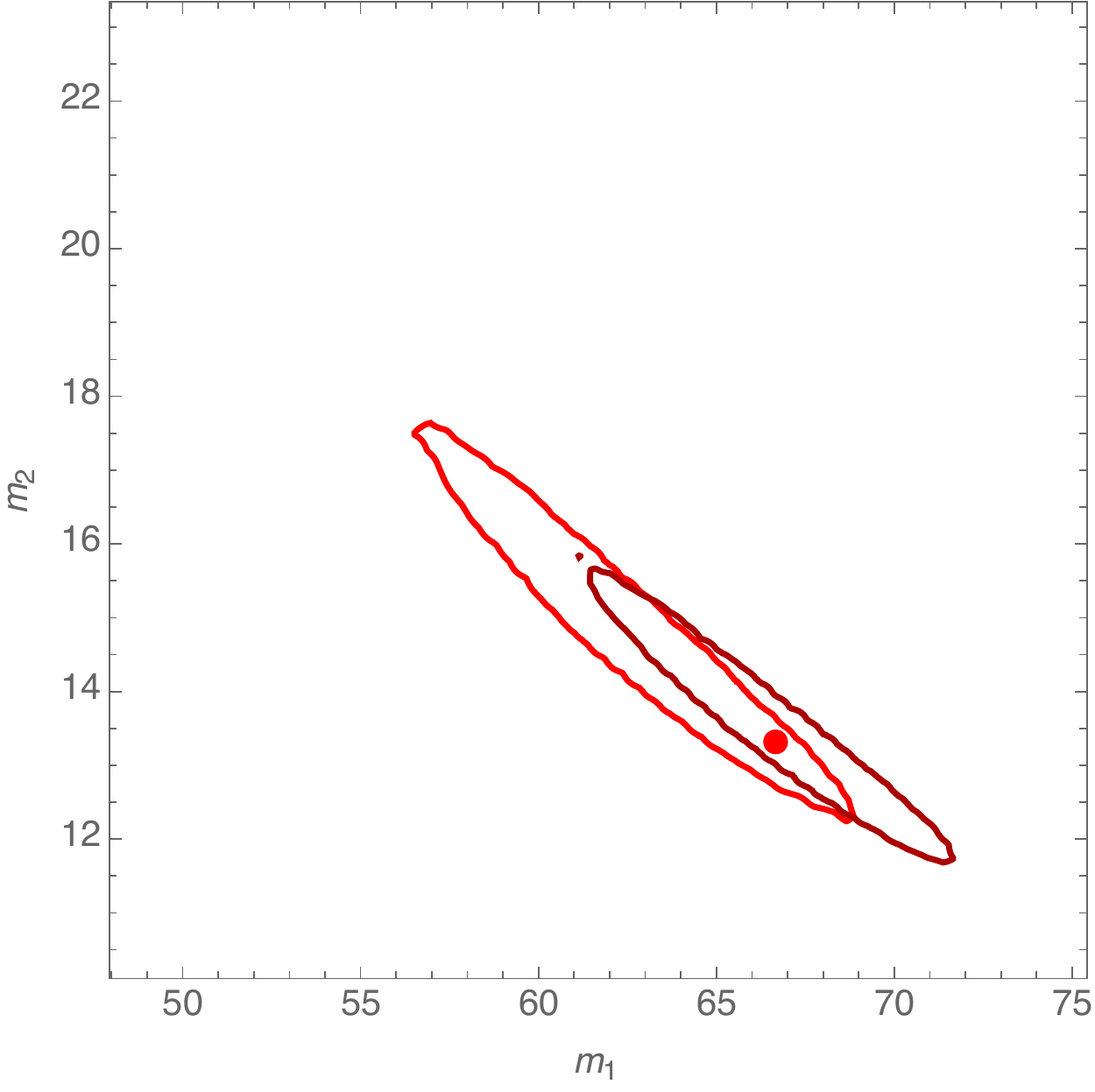}
\includegraphics[width=\columnwidth]{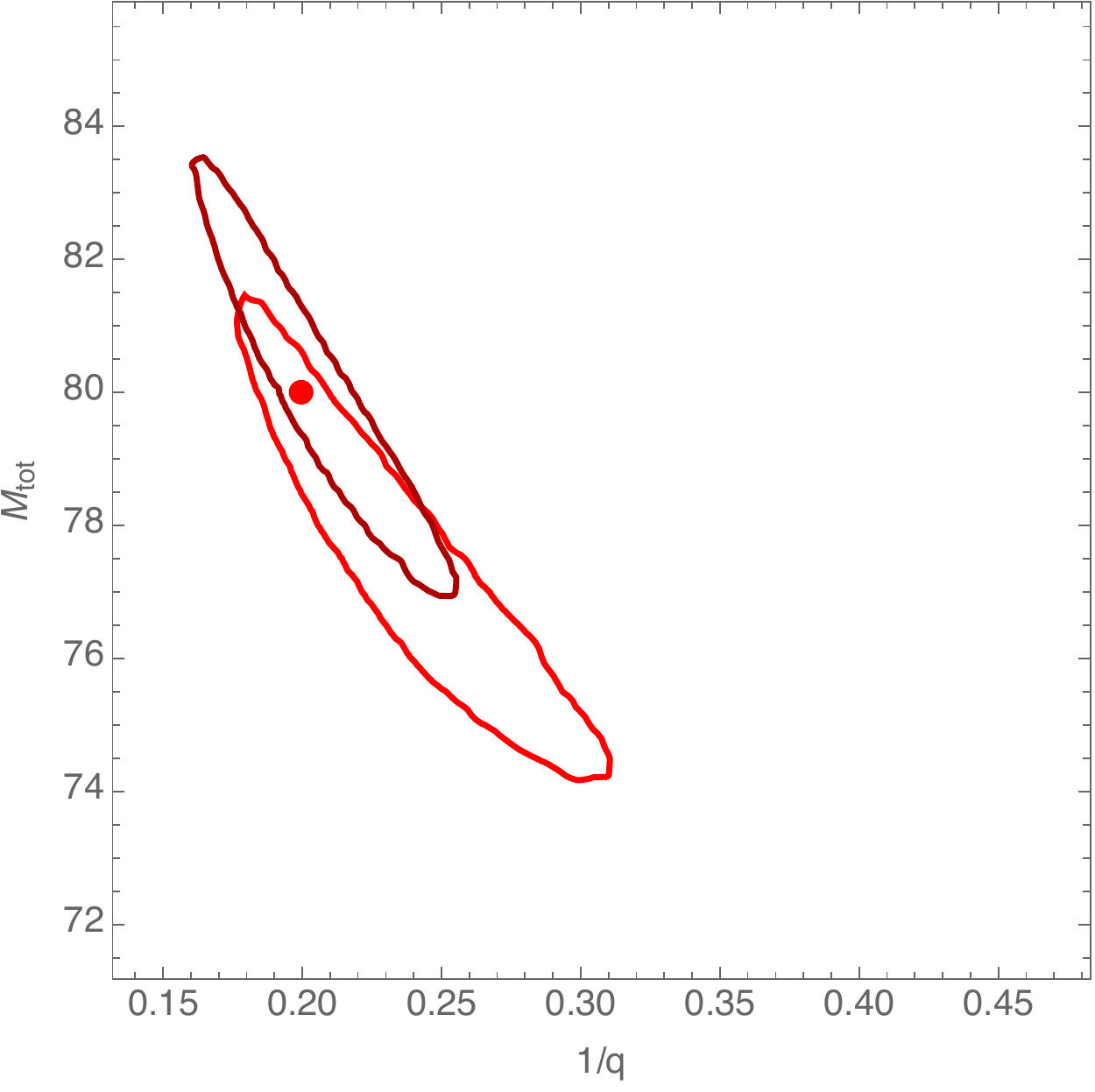}
\caption{\label{fig:NoBiasPE:HigherModeExample:NoSpinTest}\textbf{Parameter recovery with and without higher modes (assuming zero spin)}:
Red dot shows the parameters of a synthetic nonprecessing binary, generated as in Figure
    \ref{fig:NoBiasPE:HigherModeExample}.   The dark red contour shows inference using \textsc{EOBNRv2HM} (a nonspinning model
    including higher modes); the light red contour shows parameter inferences drawn using \textsc{SEOBNRv4}, assuming both black
holes have zero spin.   The former region is smaller than the latter, and more closely centered on the true parameters.  This figure illustrates the previously-appreciated fact that inference including higher modes 
draws sharper conclusions with smaller biases, using the examples previously used in this work.
}
\end{figure}

In principle, inference with higher modes and precession can more efficiently extract information  from and produce significantly narrower
posteriors for BH-BH mergers; see, e.g.,~\cite{2015PhRvD..92b2002G,2017CQGra..34n4002O} as well as our Figure \ref{fig:NoBiasPE:HigherModeExample:NoSpinTest}.   Proof-of-concept new models
containing these modes for precessing binaries  have only recently been introduced
\cite{2017PhRvD..95j4023B,2017arXiv170507089B}, and have not yet been extensively applied to parameter inference.  

\skipme{
\subsection{Model inference and bias}
The biases illustrated in their investigation, if not accounted for, will have their most pernicious inference in model
selection and inference.

}

\section{Conclusions}
Using concrete binary black hole parameters consistent with LIGO's observations to date, we have demonstrated that the
two models used to infer BH-BH binary parameters can often be significantly inconsistent with one another, as measured
by their overlap.  Differences are most significant for parameters corresponding to strongly precessing BH-BH binaries, viewed from directions where
modulations from  precession are strongly imprinted on the outgoing radiation.  
Again using concrete and unexceptional binary parameters -- at the signal strengths,  masses, and spins corresponding to
current observations  --  we demonstrate that these model differences are more than
sufficient to significantly bias parameter inference for astrophysically interesting quantities, like the joint distribution of the
most massive BH's mass and spin.    
In principle, these systematic differences could be identified by parameter inference performed with both models,
identifying regions of disagreement.  In practice,  however, for long  BH-BH merger signals like GW151226, the
computational cost of large-scale parameter estimation  with \SEOB and conventional parameter inference tools remains
cost-prohibitive at present.   

After the discovery of GW150914, \citet{LIGO-O1-PENR-Systematics} performed a systematic parameter investigation study,
assessing how reliably the (known) parameters of synthetic signals were recovered.  That investigation used  \IMRP
for parameter inference; full numerical relativity simulations as sources; and emphasized source parameters similar to
GW150914.      Our study complements this initial investigation by directly comparing the two models used for inference;
by using both model- and NR-based synthetic sources; and by using source parameters consistent with subsequent LIGO
observations. 

At present, the posterior distributions for any individual astrophysically interesting parameter is often large, due to
a combination of modest signal strength, brevity, and some degree of model incompleteness.    In particular, even for
the most extreme examples of synthetic inference studied here, where model disagreements were most substantial, the
one-dimensional posterior probability distributions still contained the known value.    While these biases described in
this work are not always large compared to the posterior's extent, these biases could complicate attempts to use
multiple events to draw astrophysical inferences about compact binary populations. 
For the immediate future, parameter inference for BH-BH binaries should be performed with  multiple models (including numerical
relativity), and  carefully validated by performing inference under controlled circumstances with similar synthetic
events.  Extensive followup studies of this kind were performed for GW150914~\cite{PEPaper,NRPaper,LIGO-O1-PENR-Systematics} and  GW170104~\cite{2017PhRvL.118v1101A,NRFollowupPaperGW170104}, so their  general
parameters are not in doubt.

\begin{acknowledgements}
    ARW acknowledges support from the RIT's Office of the Vice President for Research through the FGWA SIRA initiative.
    ROS is supported by NSF AST-1412449, PHY-1505629, and PHY-1607520.
    JAC and JCB are supported by NSF PHY-1505824, PHY-1505524, and PHY-1333360.
    PK gratefully acknowledges support for this research at CITA from NSERC of Canada, the Ontario Early Researcher Awards Program, the Canada Research Chairs Program, and the Canadian Institute for Advanced Research.
    JV was supported by the UK STFC grant ST/K005014/1.
    ROS  acknowledges the hospitality of the Aspen Center for Physics, supported by NSF PHY-1607611, where this work was completed.  
The authors thank Katerina Chatziioannou for comments on the manuscript.
We note that after our limited  investigations on the effect of priors on GW170104 was performed, a thorough study on the effect of priors was performed by Vitale et al
\cite{2017arXiv170704637V}.  
The authors thank to the LIGO Scientific Collaboration for access to the data and gratefully acknowledge the support of the United States National Science Foundation (NSF) for the construction and operation of the LIGO Laboratory and Advanced LIGO as well as the Science and Technology Facilities Council (STFC) of the United Kingdom, and the Max-Planck-Society (MPS) for support of the construction of Advanced LIGO. Additional support for Advanced LIGO was provided by the Australian Research Council.

\end{acknowledgements}

\bibliography{bibads}
\end{document}